\documentclass[nopreprint,twocolumn,author-year]{jasatex}

\usepackage{epsfig, epsf, graphicx, subfigure} 
\usepackage{bm,mathrsfs,xspace}           
\usepackage{amsmath,amsfonts,amssymb,amsthm,amscd}      
\usepackage{algorithm, algorithmic}                     
\usepackage{array,dcolumn}
\usepackage{hyperref}
\usepackage[safe]{tipa}
\usepackage[usenames,dvipsnames]{color}
\usepackage{enumitem} 

\hypersetup{
    breaklinks=true,
    colorlinks=true,        
    linkcolor=black,        
    citecolor=black,         
    filecolor=black,         
    urlcolor=black           
}

\begin{document}

\title[
Kalman-based formant and antiformant tracking
]{%
{KARMA}: Kalman-based autoregressive moving average modeling and inference for formant and antiformant tracking
}
\thanks{Portions of this work were presented at the {INTERSPEECH} conference in Antwerp, Belgium, in August 2007 \citep{Rudoy2007}.}

\author{Daryush D. Mehta}
\email{dmehta@seas.harvard.edu}
\altaffiliation[also with ]{the Center for Laryngeal Surgery and Voice Rehabilitation, Massachusetts General Hospital, Boston, Massachusetts 02114}

\author{Daniel Rudoy}
\author{Patrick J. Wolfe}
\altaffiliation[Also with ]{the Department of Statistics, Harvard University, and the Speech and Hearing Bioscience and Technology Program, Harvard-MIT
Division of Health Sciences and Technology, Massachusetts Institute of Technology, Cambridge, Massachusetts 02139.}

\affiliation{School of Engineering and Applied Sciences, Harvard University, Cambridge, Massachusetts 02138}

\date{\today}

\begin{abstract}
Vocal tract resonance characteristics in acoustic speech signals are classically tracked using frame-by-frame point estimates of formant frequencies followed by candidate selection and smoothing using dynamic programming methods that minimize \emph{ad hoc} cost functions. The goal of the current work is to provide both point estimates and associated uncertainties of center frequencies and bandwidths in a statistically principled state-space framework. Extended Kalman (K) algorithms take advantage of a linearized mapping to infer formant and antiformant parameters from frame-based estimates of autoregressive moving average (ARMA) cepstral coefficients. Error analysis of KARMA, WaveSurfer, and Praat is accomplished in the all-pole case using a manually marked formant database and synthesized speech waveforms. KARMA formant tracks exhibit lower overall root-mean-square error relative to the two benchmark algorithms, with third formant tracking more challenging. Antiformant tracking performance of KARMA is illustrated using synthesized and spoken nasal phonemes. The simultaneous tracking of uncertainty levels enables practitioners to recognize time-varying confidence in parameters of interest and adjust algorithmic settings accordingly.

\end{abstract}



\maketitle

\section{Introduction}

Speech formant tracking has received continued attention over the past sixty years to better characterize formant motion during vowels as well as vowel-consonant boundaries. The de facto approach to resonance estimation involves waveform segmentation and the assumption of an all-pole model characterized by second-order digital resonators \citep{Schafer1970}. The center frequency and bandwidth of each resonator are then estimated through picking peaks in the all-pole spectrum or finding roots of the prediction polynomial. Tracking these estimates across frames is typically accomplished via dynamic programming methods that minimize cost functions to produce smoothly-varying trajectories.

This general formant-tracking algorithm is implemented in WaveSurfer \citep{Wavesurfer185} and Praat \citep{Praat}, speech analysis tools that enjoy widespread use in the speech recognition, clinical, and linguistic communities. There are, however, numerous shortcomings to this classical approach. For example, formant track smoothing and correction (e.g., for large frequency jumps) are performed in an \emph{ad hoc} manner that precludes that ability to apply statistical analysis to obtain confidence intervals around the estimated tracks.

Initial development of the formant tracking approach described here has been reported by \cite{Rudoy2007} using a manually marked formant database for error analysis \citep{Deng2006a}. The current work continues this line of research and offers two main contributions. The first provides improvements to the Kalman-based autoregressive approach of \cite{Deng2007} and extensions to enable antiformant frequency and bandwidth tracking in a Kalman-based autoregressive moving average (KARMA) framework. The second empirically determines the performance of the KARMA approach through visual and quantitative error analysis and compares this performance with that of WaveSurfer and Praat.

\subsection{Classical formant tracking algorithms}

Linear predictive coding (LPC) models have been shown to efficiently encode source/filter characteristics of the acoustic speech signal \citep{Atal1971}. To extract frame-by-frame formant parameters, the poles of the LPC spectrum can be computed as the roots of the prediction polynomial, peaks in the LPC spectrum, or peaks in the second derivative of the frequency spectrum \citep{Christensen1976}. The first complete formant \emph{tracker} over multiple continuous speech frames incorporated spectral peak-picking, selection of formants from the candidate peaks using continuity constraints, and voicing detection to handle silent and unvoiced speech segments \citep{McCandless1974}. Extensions to LPC analysis incorporate autoregressive moving average (ARMA) models that added estimates of candidate zeros associated with anti-resonances during consonantal and nasal speech sounds \citep{Steiglitz1977,Atal1978}.

\begin{figure*}[t!]
  \includegraphics{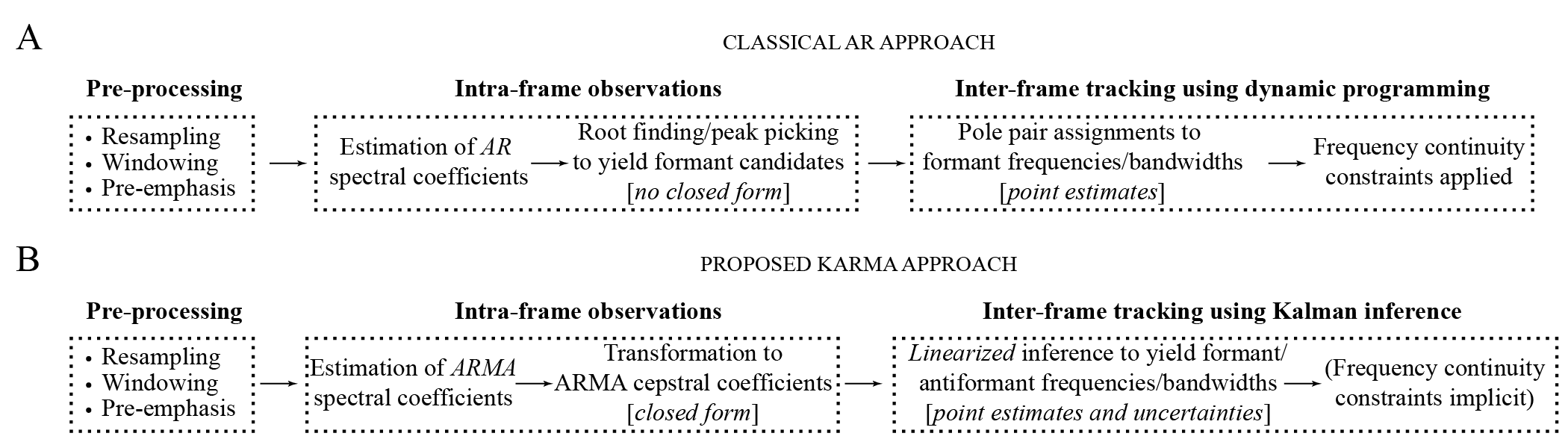}
  \caption{Illustration of the (A) classical and (B) proposed approaches to formant tracking. Key advantages to the proposed KARMA approach include intra-frame observation of autoregressive moving average (ARMA) parameters for both formant and antiformant tracking, inter-frame tracking using linearized Kalman (K) inference, and the availability of both point estimates and uncertainties for each trajectory.}
  \label{fig:ftApproaches}
\end{figure*}

Fig.~\ref{fig:ftApproaches}A illustrates the classical tracking process for the all-pole case. Following pre-processing steps, LPC spectral coefficients yield intra-frame point estimates of candidate frequency and bandwidth parameters via root finding or peak-picking. Intra-frame parameter estimation can be accomplished using a number of methods \citep{Atal1971,Atal1978,Broad1989,Yegnanarayana1978}, and inter-frame parameter selection and smoothing can be performed by minimizing various cost functions in a dynamic programming environment \citep{Wavesurfer185,Praat}. Note that the required root-finding (or peak-picking) procedure cannot be written in closed form. Consequently, statistical analysis (distributions, bias, variance) of the resultant formant and bandwidth estimates is challenging. Alternative spectrographic representations primarily apply to sustained vowels and require significant manual interaction \citep{Fulop2010}.

\subsection{Statistical formant tracking algorithms}

Probabilistic and statistical models for tracking formants have gained widespread use in the past $15$ years with motivation from automatic speech recognition applications. The first such probabilistic model was introduced by \cite{Kopec1986}, in which a hidden Markov model was used to constrain the evolution of vector-quantized sets of formant frequencies and bandwidths. Similarly, a state-space dynamical model can appropriately constrain the evolution of formant parameters, where observations of the acoustic speech waveform are linked through nonlinear relationships to ``hidden'' states (formant parameters) that evolve over time. Inference of the state values can be performed by variants of Kalman filter algorithms \citep{Kalman1960}.

In these algorithms, \emph{ad hoc} assignment of poles and zeros to appropriate formant indices is precluded by the inherent association of spectral/cepstral coefficients to formant and antiformant frequencies and bandwidths. The first reported state-space approach to formant tracking inferred formant frequencies and bandwidths directly from LPC spectral coefficients \citep{Rigoll1986}. An extension to this LPC approach was made by \cite{Toyoshima1991} to build a tracker that inferred frequencies and bandwidths of both formants and antiformants from time-varying ARMA spectral coefficients \citep{Miyanaga1986}. More recent state-space models define the observations as coefficients in the LPC \emph{cepstral} domain \citep{Zheng2004,Deng2007}, providing statistical methods to support or refute empirical relations obtained between low-order cepstral coefficients and formant frequencies \citep{Broad1989}.

The proposed KARMA (Kalman-based autoregressive moving average) approach explores the performance of such a state-space model with {ARMA} cepstral coefficients as observations to track formant and antiformant parameters. Taking advantage of a linearized mapping between frequency and bandwidth values and cepstral coefficients, KARMA applies Kalman inference to yield point estimates and uncertainties for the output trajectories.

\section{Methods}
\label{sec:methods}

Fig.~\ref{fig:ftApproaches}B illustrates the proposed statistical modeling approach to formant and antiformant tracking. This approach affords several advantages over classical approaches: (1)~both formant and antiformant trajectories are tracked, (2)~both frequency and bandwidth estimates are propagated as distributions instead of point estimates to provide for uncertainty quantification, and (3)~pole/zero assignment to formants/antiformants is made through a linearized cepstral mapping instead of candidate selection using \emph{ad hoc} cost functions.

\subsection{Step 1: Pre-processing}
\label{sec:Preprocessing}

The sampled acoustic speech waveform $s[m]$ is first windowed into short-time frames $s_t[m] = s[m]w_t[m]$ using overlapping windows $w_t[m]$ with frame index $t$. Each short-time frame $s_t[m]$ is then pre-emphasized via
\begin{equation}
		\label{eq:preemphasis}
		s_t[m] = s_t[m] - \gamma s_t[m-1],
\end{equation}
where $\gamma$ is the pre-emphasis coefficient defining the inherent high-pass filter characteristic that is typically applied to equalize energy across the speech spectrum for improved model fitting.

\subsection{Step 2: Intra-frame observation generation}
\label{sec:intraframeObs}

\subsubsection{ARMA model of speech}
\label{sec:armaModel}

Following windowing and pre-emphasis, the acoustic waveform $s_t[m]$ is modeled as a stochastic ARMA$(p,q)$ process:
\begin{equation}
		\label{eq:ARMAprocess}
		s_t[m] = \sum_{i=1}^{p}a_is_t[m-i] + \sum_{j=1}^{q}b_ju[m-j] + u[m],
\end{equation}
where $a_i$ are the $p$ AR coefficients, $b_j$ are the $q$ MA coefficients, and $u[m]$ is the stochastic excitation waveform. The $z$-domain transfer function associated with Eq.~\eqref{eq:ARMAprocess} is
\begin{equation}
		\label{eq:armaSpectrum}
    T(z) \triangleq \frac{1+\sum_{j=1}^{q}{b}_jz^{-j}}{1-\sum_{i=1}^{p}{a}_iz^{-i}}.
\end{equation}
A number of standard spectral estimation techniques can be employed in order to fit data to the ARMA$(p,q)$ model \citep[see][for a recent review of ARMA estimation methods]{Marelli2010}. In the current study, ARMA estimation was performed using the `armax' function in MATLAB's System Identification toolbox (The MathWorks, Natick, MA), which implements an iterative method to minimize a quadratic error prediction criterion \citep[][Section 10.2]{Ljung1999}.

\subsubsection{Generation of observations: ARMA cepstral coefficients}
\label{sec:arma2c}

In the proposed approach, the ARMA spectral coefficients in Eq.~\eqref{eq:armaSpectrum} are transformed to the complex cepstrum before inferring formant characteristics. This mapping from ARMA spectral coefficients to ARMA cepstral coefficients has been derived in the all-pole case \citep[e.g.,][]{Deng2006} and can be extended to account for the presence of zeros in the spectrum. Letting ${C}_n$ denote the $n$th cepstral coefficient,
\begin{equation}
\label{eq:cpcz}
	{C}_n = {c}_n - {c}'_n,
\end{equation}
where $C_n$ depends on separate contributions from the denominator and numerator of the ARMA model through the following recursive relationships:
\begin{subequations}
\label{eq:arma2lpcc}
    \begin{equation}
    \label{eq:arma2lpccP}
		    c_n = \begin{cases}
		        a_n & \text{if} \quad n = 1 \\
		        a_n + \sum_{i=1}^{n-1} \left(\frac{i}{n}\right) a_{n-i} c_{i} & \text{if} \quad 1 < n \leq p \\
		        \sum_{i=n-p}^{n-1} \left(\frac{i}{n}\right) a_{n-i} c_i  & \text{if} \quad p < n \text{,}
		    \end{cases}
    \end{equation}
    \begin{equation}
    \label{eq:arma2lpccZ}
		    c'_n = \begin{cases}
		        b_n & \text{if} \quad n = 1 \\
		        b_n + \sum_{j=1}^{n-1} \left(\frac{j}{n}\right) b_{n-j} c'_j & \text{if} \quad 1 < n \leq q \\
		        \sum_{j=n-q}^{n-1} \left(\frac{j}{n}\right) b_{n-j} c'_j  & \text{if} \quad q < n \text{.}
		    \end{cases}
    \end{equation}
\end{subequations}
Derivation of Eqs.~\eqref{eq:arma2lpcc} is given in the Appendix. The proof is derived under the minimum-phase assumption that constrains the poles and zeros of the ARMA transfer function to lie within the unit circle.

\subsection{Step 3: Inter-frame parameter tracking}

The proposed algorithm tracks point estimates and uncertainties for $I$ formants and $J$ antiformants from frame to frame. To accommodate the temporal dimension, the parameters of frame $t$ are placed in column vector $\bm{x}_t$:
\begin{equation}
\label{eq:stateVector}
    \bm{x}_t \triangleq
    \begin{pmatrix}
      {f}_1 \ldots {f}_I & {b}_1 \ldots {b}_I & {f}'_1 \ldots {f}'_J & {b}'_1 \ldots {b}'_J
    \end{pmatrix}^T \text{,}
\end{equation}
where $({f}_i,{b}_i)$ is the frequency/bandwidth pair of the $i$th formant and $({f}'_j,{b}'_j)$ is the frequency/bandwidth pair for the $j$th antiformant.

\subsubsection{Observation model}
\label{sec:fb2c}

Inference of the output parameters is facilitated by a closed-form mapping from the state vector $\bm{x}_t$ to the observed cepstral coefficients $C_n$ in Eq.~\eqref{eq:cpcz}. Extending the speech production model of \cite{Schafer1970} to capture zeros, we assume that the transfer function $T(z)$ of the ARMA model can be written as a cascade of $I$ second-order digital resonators and $J$ second-order digital anti-resonators:
\begin{equation}
    \label{eq:envelopeSpectrum}
    T(z) = \frac{\prod_{j = 1}^{J} (1 - {\beta}_jz^{-1})(1 - \overline{{\beta}_j}z^{-1})}{\prod_{i = 1}^{I} (1 - {\alpha}_iz^{-1})(1 - \overline{{\alpha}_i}z^{-1})} \text{,}
\end{equation}
where $({\alpha}_i, \overline{\alpha}_i)$ and $({\beta}_j, \overline{\beta}_j)$ denote complex-conjugate pole and zero pairs, respectively. Each pole and zero are parameterized by a center frequency and 3-dB bandwidth (both in units of Hertz) using the following relations:
\begin{subequations}
\label{eq:fb2pz}
\begin{align}
    ({\alpha}_i, \overline{\alpha}_i) &= \exp \left ( \frac{-\pi {b}_i  \pm 2\pi \sqrt{-1} {f}_i} {f_s} \right ) \text{,} \\
    ({\beta}_j,  \overline{\beta}_j)  &= \exp \left ( \frac{-\pi {b}'_j \pm 2\pi \sqrt{-1} {f}'_j}{f_s} \right ) \text{,}
\end{align}
\end{subequations}
where $f_s$ is the sampling rate (in Hz).

Performing a Taylor-series expansion of $\log T(z)$ yields
\begin{equation}
\label{eq:arma2ceplong}
		\begin{split}
    \log T(z) = \sum_{i=1}^{I} \sum_{n=1}^\infty \frac{\left( {\alpha}_i^n + \overline{\alpha}_i^n \right)}{n}z^{-n} - \\
    \sum_{j=1}^{J} \sum_{n=1}^\infty \frac{\left( {\beta}_j^n + \overline{{\beta}_j}^n \right)}{n}z^{-n}.
    \end{split}
\end{equation}
Recalling that ${C}_n$ is the $n$th cepstral coefficient, $\log T(z) = {C}_0 + \sum_{n=1}^\infty {C}_n z^{-n}$. Thus, equating the coefficients of powers of $z^{-1}$ leads to
\begin{equation}
\label{eq:arma2cepshort}
    {C}_n = \frac{1}{n} \sum_{i=1}^{I} \left( {\alpha}_i^n + \overline{{\alpha}_i}^n \right ) -
    		\frac{1}{n} \sum_{j=1}^{J} \left( {\beta}_j^n + \overline{{\beta}_j}^n \right) \text{.} \\
\end{equation}
Finally, inserting ${\alpha}_i$ and ${\beta}_j$ from Eqs.~\eqref{eq:fb2pz} into Eq.~\eqref{eq:arma2cepshort} yields the following observation model $h(\bm{x}_t)$ that maps elements of $\bm{x}_t$ to $C_n$:
\begin{equation}
    \label{eq:fb2cp}
	    \begin{split}
    h(\bm{x}_t) \triangleq &{C}_n \\
          = &\frac{2}{n}\sum_{i=1}^{I} \exp \left (- \frac{\pi n}{f_s}{b}_i\right )\cos \left ( \frac{2\pi n }{f_s}{f}_i \right ) - \\
    &\frac{2}{n}\sum_{j=1}^{J} \exp \left (- \frac{\pi n}{f_s}{b}'_j\right )\cos \left ( \frac{2\pi n }{f_s}{f}'_j \right ) \text{.}
		\end{split}
\end{equation}

\subsubsection{State-space model}
\label{sec:stateSpace}

We adopt a state-space framework similar to that by \cite{Deng2007} to model the evolution of the state vector in Eq.~\eqref{eq:stateVector} from frame $t$ to frame $t+1$:
\begin{subequations}
\label{eq:stateSpace}
\begin{align}
      \bm{x}_{t+1} &= \bm{F} \bm{x}_t + \bm{w}_t, \\
      \bm{y}_t     &= h \left (\bm{x}_t \right ) + \bm{v}_t \text{,}
\end{align}
\end{subequations}
where $\bm{F}$ is the state transition matrix, and $\bm{w}_t$ and $\bm{v}_t$ are uncorrelated white Gaussian sequences with covariance matrices $\bm{Q}$ and $\bm{R}$, respectively. The function $h(\bm{x}_t)$ is the nonlinear mapping of Eq.~\eqref{eq:fb2cp}, and vector $\bm{y}_t$ consists of \emph{estimates} of the first $N$ cepstral coefficients of $C_n$ (not including the zeroth coefficient). The initial state $\bm{x}_0$ follows a normal distribution with mean $\bm{\mu}_0$ and covariance $\bm{\Sigma}_0$. The state-space model of Eqs.~\eqref{eq:stateSpace} is thus parameterized by the set $\bm{\theta}$:
\begin{equation}
    \label{eq:modelParams}
    \bm{\theta} \triangleq \left ( \bm{F}, \bm{Q}, \bm{R}, \bm{\mu}_0, \bm{\Sigma}_0 \right ) \text{.}
\end{equation}

\subsubsection{Linearization via {T}aylor approximation}
\label{sec:linearization}

The cepstral mapping in Eq.~\eqref{eq:fb2cp} can be linearized to enable \emph{approximate} minimum-mean-square-error (MMSE) estimates of the tracked states via the extended Kalman filter. The mapping $h(\bm{x}_t)$ is linearized by computing the first-order terms of the Taylor-series expansion of $C_n$ in Eq.~\eqref{eq:fb2cp}:
\begin{align*}
        \frac{\partial {C}_n}{\partial {f}_i } &= -\frac{4\pi}{f_s} \exp \left (- \frac{\pi n}{f_s}{b}_i \right )\sin \left ( \frac{2\pi n }{f_s}{f}_i  \right ),\\
        \frac{\partial {C}_n}{\partial {b}_i } &= -\frac{2\pi}{f_s} \exp \left (- \frac{\pi n}{f_s}{b}_i \right )\cos \left ( \frac{2\pi n }{f_s}{f}_i  \right ),\\
        \frac{\partial {C}_n}{\partial {f}'_j} &=  \frac{4\pi}{f_s} \exp \left (- \frac{\pi n}{f_s}{b}'_j\right )\sin \left ( \frac{2\pi n }{f_s}{f}'_j \right ),\\
        \frac{\partial {C}_n}{\partial {b}'_j} &=  \frac{2\pi}{f_s} \exp \left (- \frac{\pi n}{f_s}{b}'_j\right )\cos \left ( \frac{2\pi n }{f_s}{f}'_j \right ).
\end{align*}
The Jacobian matrix $\bm{H}_t$ thus consists of four sub-matrices for each frame $t$:
\begin{equation}
\label{eq:Ht}
    \bm{H}_t \triangleq \begin{pmatrix} \bm{H}({f}_i) & \bm{H}({b}_i) & \bm{H}({f}'_j) & \bm{H}({b}'_j) \end{pmatrix} \text{,}
\end{equation}
where $\bm{H}({f}_i)$ and $\bm{H}({b}_i)$ each consists of $N$ rows and $p/2$ columns:
\begin{subequations}
    \begin{align}
       \bm{H}({f}_i) &\triangleq
       \begin{pmatrix}
            \frac{\partial {C}_1}{\partial {f}_1} & \frac{\partial {C}_1}{\partial {f}_2} &\cdots & \frac{\partial {C}_1}{\partial {f}_{p/2}} \\
            \frac{\partial {C}_2}{\partial {f}_1} & \frac{\partial {C}_2}{\partial {f}_2} &\cdots & \frac{\partial {C}_2}{\partial {f}_{p/2}} \\
            \vdots & \vdots & \ddots & \vdots \\
            \frac{\partial {C}_N}{\partial {f}_1} & \frac{\partial {C}_N}{\partial {f}_2} &\cdots & \frac{\partial {C}_N}{\partial {f}_{p/2}}
       \end{pmatrix} \text{,} \\
       \bm{H}({b}_i) &\triangleq
       \begin{pmatrix}
            \frac{\partial {C}_1}{\partial {b}_1} & \frac{\partial {C}_1}{\partial {b}_2} &\cdots & \frac{\partial {C}_1}{\partial {b}_{p/2}} \\
            \frac{\partial {C}_2}{\partial {b}_1} & \frac{\partial {C}_2}{\partial {b}_2} &\cdots & \frac{\partial {C}_2}{\partial {b}_{p/2}} \\
            \vdots & \vdots & \ddots & \vdots \\
            \frac{\partial {C}_N}{\partial {b}_1} & \frac{\partial {C}_N}{\partial {b}_2} &\cdots & \frac{\partial {C}_N}{\partial {b}_{p/2}}
        \end{pmatrix} \text{,}
    \end{align}
\end{subequations}
and $\bm{H}({f}'_j)$ and $\bm{H}({b}'_j)$ are defined analogously each with $N$ rows and $q/2$ columns.

\subsubsection{Kalman-based inference}

Given observations $\bm{y}_t$ for frame indices $1$ to $T$, the extended Kalman smoother (EKS) can be used to compute the mean $\bm{m}_{t|T}$ (point estimates) and covariance $\bm{P}_{t|T}$ (estimate uncertainties) of each parameter in $\bm{x}_t$. Table~\ref{tab:KalmanRecursions} displays the steps of the EKS, which employs a two-pass filtering (forward) and smoothing (backward) procedure. For real-time processing, the forward filtering stage may be applied without a backward smoothing procedure; naturally, this will lead to larger uncertainties in the corresponding parameter estimates.

\begin{table}
\caption{Extended Kalman smoother algorithm.}
\label{tab:KalmanRecursions}
\begin{ruledtabular}
\begin{tabular}{p{\columnwidth}}
    \begin{enumerate}
     \item Initialization: Set $\bm{m}_{0|0} = \bm{\mu}_0$ and $\bm{P}_{0|0} = \bm{\Sigma}_0$
     \item Filtering: Repeat for $t = 1, \ldots, T$
				 \begin{eqnarray}
            \quad \quad \quad \quad \bm{m}_{t|t-1} &=& \bm{F}\bm{m}_{t-1|t-1} \nonumber \\
            \quad \quad \quad \quad \bm{P}_{t|t-1} &=& \bm{F}\bm{P}_{t-1|t-1}\bm{F}^T + \bm{Q} \nonumber \\
            \quad \quad \quad \quad \bm{K}_t       &=& \bm{P}_{t|t-1}\bm{H}_{t}^T \left (\bm{H}_{t}\bm{P}_{t|t-1}\bm{H}_{t}^T + \bm{R} \right )^{-1} \label{eq:kalmanGain} \\
            \quad \quad \quad \quad \bm{m}_{t|t}   &=& \bm{m}_{t|t-1} + \bm{K}_t(\bm{y}_t - h(\bm{m}_{t|t-1})) \nonumber \\
            \quad \quad \quad \quad \bm{P}_{t|t}   &=& \bm{P}_{t|t-1} - \bm{K}_t \bm{H}_{t} \bm{P}_{t|t-1} \nonumber
         \end{eqnarray}
     \item Smoothing: Repeat for $t = T, \ldots, 1$
     		 \begin{eqnarray*}
         		\quad \quad \quad \quad \bm{S}_{t}     &=& \bm{P}_{t-1|t-1}\bm{F}^T\bm{P}_{t|t-1}^{-1} \\
            \quad \quad \quad \quad \bm{m}_{t-1|T} &=& \bm{m}_{t-1|t-1} + \bm{S}_{t} \left (\bm{m}_{t|T} - \bm{F} \bm{m}_{t-1|t-1}\right )\\
            \quad \quad \quad \quad \bm{P}_{t-1|T} &=& \bm{P}_{t-1|t-1} + \bm{S}_{t} \left ( \bm{P}_{t|T} - \bm{P}_{t-1|t-1} \right )\bm{S}^T_{t}
         \end{eqnarray*}
    \end{enumerate}
\end{tabular}
\end{ruledtabular}
\end{table}

Care must be taken when approximating the observation model of Eq.~\eqref{eq:fb2cp} to avoid suboptimal performance or algorithm divergence in the case of the Kalman filter \citep{Julier1997}. To verify the appropriateness of the linearization in Section~\ref{sec:linearization} in this setting, comparisons are made to a more computationally intensive method of stochastic computation termed a particle filter, which approximates the densities in question by sequentially propagating a fixed number of samples, or ``particles,'' and hence avoids the linearization of Eq.~\eqref{eq:fb2cp}.

\begin{figure}
  \includegraphics{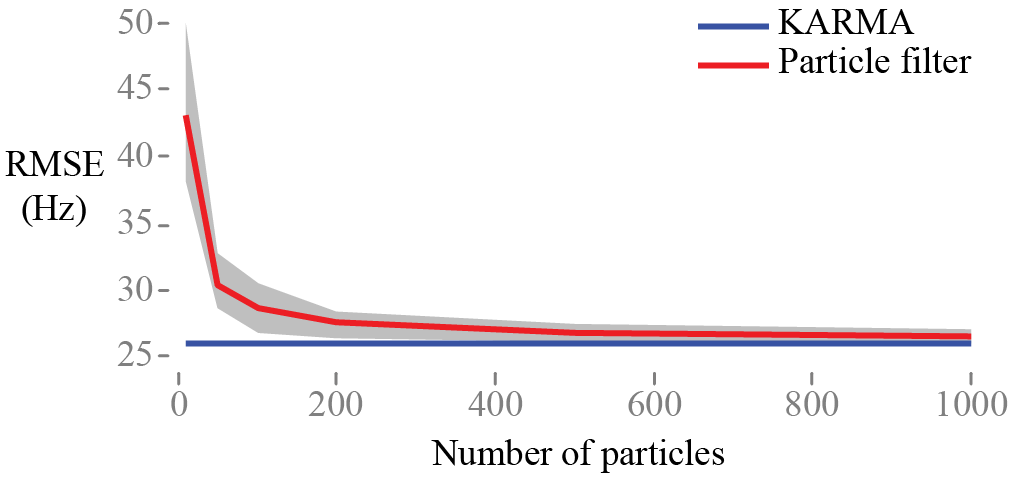}
  \caption{(Color online) Comparison of KARMA (blue) and particle filter (red) tracking performance in terms of root-mean-square error (RMSE) averaged over 25 Monte Carlo trials and reported with 95~\% confidence intervals (gray).}
  \label{fig:ft_EKFvsPF}
\end{figure}

Twenty-five Monte Carlo simulations of $100$-sample data sequences were performed according to Eqs.~\eqref{eq:stateSpace} with four complex-conjugate pole pairs ($p=8$, $q=0$) and $N=15$ cepstral coefficients. Fig.~\ref{fig:ft_EKFvsPF} compares the output of KARMA using the extended Kalman filter of Table~\ref{tab:KalmanRecursions} and that of a particle filter in terms of root-mean-square error (RMSE) as a function of the number of particles, averaged over all formant frequency tracks (true bandwidths were provided to both algorithms). The performance of the EKF compares favorably to that of the particle filter, even when a large number of particles is used. Similar results hold over a broad range of parameter values.

\subsubsection{Observability of states}

The model of Eq.~\eqref{eq:stateSpace} does not explicitly take into account the existence of speech and non-speech states. To continue to track or \emph{coast} parameters during silence frames, the state vector $\bm{x}_t$ can be augmented with a binary indicator variable to specify the presence of speech in the frame. The approximate MMSE state estimate is then obtained via EKS inference by modifying the Kalman gain in Eq.~\ref{eq:kalmanGain}:
\begin{equation*}
    \bm{K}_t = \bm{M}_t \bm{P}_{t|t-1}\bm{H}_{t}^T \left (\bm{H}_{t}\bm{P}_{t|t-1}\bm{H}_{t}^T + \bm{R} \right )^{-1} \text{,}
\end{equation*}
where $\bm{M}_t$ is a diagonal matrix with diagonal entries equal to $1$ or $0$ depending on the presence or absence, respectively, of speech energy in frame $t$.

In addition, to handle the presence or absence of particular tracks, the state vector $\bm{x}_t$ can be dynamically modified to include or omit corresponding frequency/bandwidth states in Eq.~\eqref{eq:stateVector}. The approximate MMSE state estimate is then obtained via EKS inference in Table~\ref{tab:KalmanRecursions} with the modified state vector. If an absent state reappears in a given frame, that state is reinitialized with corresponding entries in $\bm{\mu}_0$ and $\bm{\Sigma}_0$.

\subsubsection{Model order selection and system identification}

As is commonly done, the orders $p$ and $q$ of the ARMA model are chosen to capture as much information as possible on the peaks and valleys in the resonance spectrum, while avoiding overfitting and mistakenly capturing source-related information. The ARMA cepstral order $N$ is chosen to be at least $\max(p,q)$ so that all pole/zero information is incorporated per Eq.~\eqref{eq:arma2lpcc}. Finally, selecting $I$ and $J$ in Eq.~\eqref{eq:fb2cp} depends on the expected number of formants and anti-formants, respectively, in the speech bandwidth $f_s/2$.

Formants do not evolve independently of one another, and their temporal trajectories are not independent in frequency. In the synthesis of front vowels, it is common practice to employ a \emph{linear regression} of $f_3$ onto $f_1$ and $f_2$ \citep[e.g.]{Nearey1989}. Empirically, we found the formant cross-correlation function to decay slowly \citep{Rudoy2007}, implying that a set of formant values at frame $t$ might be helpful in predicting values of all formants at frame $t+1$. Thus, instead of setting the state transition matrix $\bm{F}$ to the identity matrix \citep{Deng2006, Deng2007}, $\bm{F}$ is estimated \emph{a priori} for a particular utterance from first-pass WaveSurfer formant frequency tracks using a linear least-squares estimator \citep{Hamilton1994}.

The state transition covariance matrix $\bm{Q}$, which dictates the frame-to-frame frequency variation, consists of a diagonal matrix with values corresponding to standard deviations of approximately $320$~Hz for center frequencies and $100$~Hz for bandwidths. These values were empirically found to follow temporal variations of speech articulation. The covariance matrix $\bm{R}$, representing the signal-to-noise ratio of the cepstral coefficients, is a diagonal matrix with elements $\bm{R}_{nn}=1/n$ for $n \in \{1, 2, \ldots , N\}$. This was observed to be in reasonable agreement with the variance of the residual vector of the cepstral coefficients derived from speech waveforms.

The center frequencies and bandwidths are initialized to $\bm{\mu}_0 = \begin{pmatrix} 500 & 1500 & 2500 & 80 & 120 & 160 \end{pmatrix}$~Hz. The initial covariance $\bm{\Sigma}_0$ is set to $\bm{Q}$.

\subsection{Summary of KARMA approach}

\begin{table}
\caption{Proposed KARMA algorithm for formant and antiformant tracking.}
\label{tab:algo}
\begin{ruledtabular}
\begin{tabular}{p{\columnwidth}}
Repeat for frames $t=1, \ldots, T$ (Online or batch mode)
    \begin{enumerate}
     	\item Pre-processing of input speech waveform $s[m]$
		     	\begin{enumerate}
		     			\item Window: $s_t[m] = s[m]w_t[m]$
		     			\item Pre-emphasize $s_t[m]$
					\end{enumerate}
			\item Intra-frame observation of $N$ cepstral coefficients
		     	\begin{enumerate}
		     			\item Estimate ARMA($p,q$) spectral coefficients $\hat{a}_i$ and $\hat{b}_j$ in Eq.~\eqref{eq:armaSpectrum}
		     			\item Convert $\hat{a}_i$ and $\hat{b}_j$ to ARMA cepstral coefficients using Eq.~\eqref{eq:cpcz}
					\end{enumerate}
			\item Inter-frame parameter tracking of $I$ formants and $J$ antiformants
		     	\begin{enumerate}
		     			\item Apply Kalman filtering step in Table~\ref{tab:KalmanRecursions}
		     			\item $\bm{m}_{t|t}$ are point estimates and diagonal elements of $\bm{P}_{t|t}$ are associated variances of the estimates
					\end{enumerate}
		\end{enumerate}
Repeat for frames $t=T, \ldots, 1$~(Batch mode only)
    \begin{enumerate}[resume]
			\item Inter-frame parameter tracking of $I$ formants and $J$ antiformants
		     	\begin{enumerate}
		     			\item Apply Kalman smoothing step in Table~\ref{tab:KalmanRecursions}
		     			\item $\bm{m}_{t|T}$ are point estimates and diagonal elements of $\bm{P}_{t|T}$ are associated variances of the estimates
					\end{enumerate}
		\end{enumerate}
\end{tabular}
\end{ruledtabular}
\end{table}

Table~\ref{tab:algo} outlines the steps of the proposed KARMA algorithm, which includes a pre-processing stage, intra-frame ARMA cepstral coefficient estimation, and inter-frame tracking of formant and antiformant parameters using Kalman inference.

\subsection{Benchmark algorithms}

Performance of KARMA is compared with that of Praat \citep{Praat} and WaveSurfer \citep{Wavesurfer185}, two software packages that see wide use among voice and speech researchers. WaveSurfer and Praat both follow the classical formant tracking approach in which frame-by-frame format frequency candidates are obtained from the all-pole spectrum and smoothed across the entire speech utterance to remove outliers and constrain the trajectories to physiologically plausibile values. Smoothing is accomplished through dynamic programming to minimize the sum of the following three cost functions: (1) the deviation between the frequency for each formant from baseline values of each frequency; (2) a measure of the quality factor ${f}_i / {b}_i$ of a formant, where higher quality factors are favored; and (3) a transition cost that penalizes large frequency jumps. The user sets weights to these cost functions to tune the algorithm's performance.

\section{Results}

Evaluation of KARMA is accomplished in the all-pole case using the vocal tract resonance (VTR) database \citep{Deng2006a}. Since the VTR database itself only yields estimates of ground truth and exhibits observable labeling errors, two speech databases are created using overlap-add of synthesis speech frames using the four VTR formant tracks. Antiformant tracking performance of KARMA is illustrated using synthesized and spoken nasal phonemes.

\subsection{Error analysis using a hand-corrected formant database}

The VTR database contains a representative subset of the TIMIT speech corpus \citep{TIMIT} that consists of $516$ diverse, phonetically-balanced utterances collated across gender, individual speakers, dialects, and phonetic contexts. The VTR database contains state information for four formant trajectory pairs (center frequency and bandwidth). The first \emph{three} center frequency trajectories were manually corrected after an initial automated pass \citep{Deng2004}. Corrections were made using knowledge-based intervention based on the speech waveform, its wideband spectrogram, word- and phoneme-level transcriptions, and phonemic boundaries.

Analysis parameters of KARMA are set to the following values: $f_s=7$~kHz, $20$~ms Hamming windows with $50$~\% overlap, $\gamma=0.7$, $p=12$ ($q=0$), and $I=3$ ($J=0$). Each frame is fit with an ARMA($12,0$) model using the autocorrelation method of linear prediction and, subsequently, transformed to $N = 15$ cepstral coefficients via Eq.~\eqref{eq:arma2lpcc}. The initial state vector is set to $\bm{x}_0 = \begin{pmatrix} 500 & 1500 & 2500 \end{pmatrix}^T$, and $\bm{\Sigma}_0$ is set to $\bm{Q}$. TIMIT phone transcriptions are used to indicate whether each frame contains speech energy or a silence region. A frame is considered silent if all its samples are labeled as a pause (\emph{pau}, \emph{epi}, \emph{h\#}), closure interval (\emph{bcl}, \emph{dcl}, \emph{gcl}, \emph{pcl}, \emph{tcl}, \emph{kcl}), or glottal stop (\emph{q}). Thus, errors due to speech activity detection are minimized, and all tracks are coasted during silent frames.

Default smoothing settings are set within WaveSurfer and Praat. Other analysis parameters are matched to KARMA: $f_s=7$~kHz, $20$~ms Hamming windows with $50$~\% overlap, $\gamma=0.7$, $p=12$ ($q=0$), and $I=3$ ($J=0$).

\begin{table}[b]
\caption[RMSE of EKS, WaveSurfer, Praat re VTR on speech frames]{\label{tab:rmse3500}Formant tracking performance of KARMA, WaveSurfer, and Praat in terms of root-mean-square error (RMSE) per formant averaged across $516$ utterances in the VTR database \citep{Deng2006a}. RMSE is only computed over speech-labeled frames.}
\begin{ruledtabular}
\begin{tabular}{cccc}
\textbf{Formant} & \textbf{KARMA} & \textbf{WaveSurfer} & \textbf{Praat} \\
\hline
$f_1$ 			& $114$~Hz & $170$~Hz & $185$~Hz \\
$f_2$ 			& $226$~Hz & $276$~Hz & $254$~Hz \\
$f_3$ 			& $320$~Hz & $383$~Hz & $303$~Hz \\
Overall 		& $220$~Hz & $276$~Hz & $247$~Hz
\end{tabular}
\end{ruledtabular}
\end{table}

Table~\ref{tab:rmse3500} summarizes the performance of KARMA, WaveSurfer, and Praat on the VTR database. The root-mean-square error (RMSE) per formant is computed over all speech frames for each utterance and then averaged, per formant, across all $516$ utterances in the database. The cepstral-based KARMA approach results in lower overall error compared to the classical algorithms, with particular gains for $f_1$ and $f_2$ tracking. Praat exhibits the lowest average error for $f_3$.

\begin{figure*}
	\includegraphics{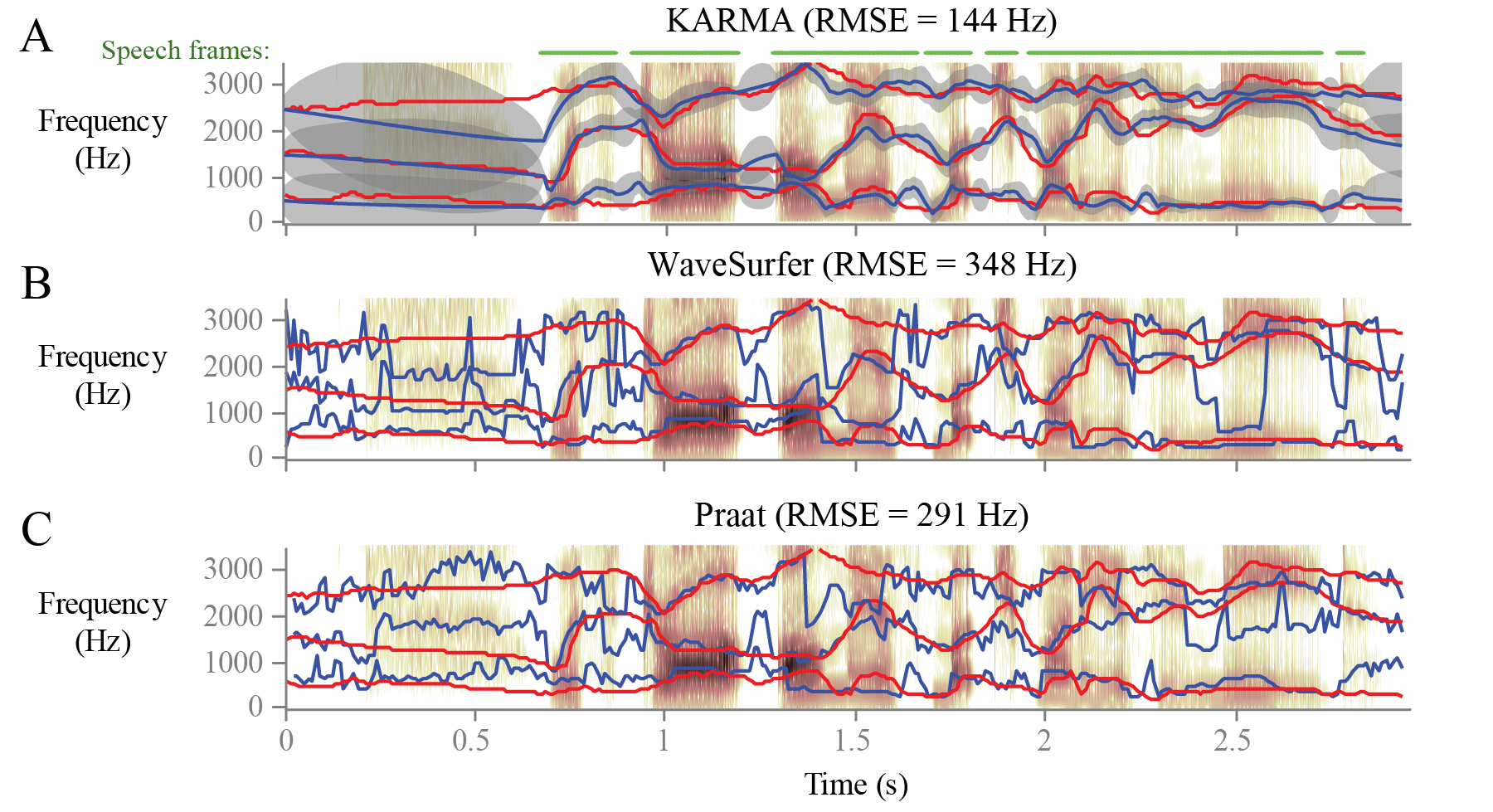}
  \caption{(Color online) Estimated formant tracks on spectrogram of VTR utterance 200: ``Withdraw only as much money as you need.'' Reference trajectories from the VTR database are shown in red along with the formant frequency tracks in blue from (A) KARMA, (B) WaveSurfer, and (C) Praat. The KARMA output additionally displays uncertainties (gray shading, $\pm1$ standard deviation) for each formant trajectory and speech-labeled frames (green). Reported root-mean-square error (RMSE) is averaged across formants conditioned on speech presence for each frame.}
  \label{fig:timit200}
\end{figure*}

Figure~\ref{fig:timit200} illustrates the formant tracks output from the three algorithms for VTR utterance $200$ spoken by an adult female. During non-speech regions (the first $700$~ms exhibits noise energy during inhalation), mean KARMA track estimates are linear with increasing uncertainty for frames that are farther from frames with formant information. Compared to the WaveSurfer and Praat tracks, KARMA trajectories are smoother and better behaved, reflecting the slow-moving nature of the speech articulators. The classical algorithms exhibit errant tracking of $f_2$ during the /i/ vowel in ``need'' at $2.5$~s that is handled by the KARMA approach.

\subsection{Error analysis using synthesized databases}

Speech waveforms in the first database (VTRsynth) are synthesized through overlap-add of frames that each follow the ARMA model of Eq.~\eqref{eq:ARMAprocess}. ARMA spectral coefficients are derived from the four formant frequency/bandwidth pairs in the corresponding frame of the VTR database utterance using the impulse-invariant transformation of a digital resonator \citep{Klatt1980}. The source excitation is white Gaussian noise during non-silence frames. Synthesis parameters are set to the following values: $f_s=16$~kHz, $20$~ms Hanning windows with $50$~\% overlap, and $p=8$ ($q=0$).

The second database (VTRsynthf0) introduces a model mismatch between synthesis and KARMA analysis by applying a Rosenberg C source waveform \citep{Rosenberg1971} instead of white noise for each frame considered voiced. The fundamental frequency of each voiced frame in the original VTR database is estimated by WaveSurfer. The VTRsynthf0 database thus includes voiced, unvoiced, and non-speech frames. Synthesis parameters are set as in the VTRsynth database. Formant trajectories from these two synthesized databases act as truer ground truth contours than in the VTR database to test the performance of the cepstral-based KARMA algorithm.

\begin{table}[b]
\caption[RMSE of EKS, WaveSurfer, Praat re VTR on speech frames]{\label{tab:rmse3500synth}Average RMSE of KARMA, WaveSurfer, and Praat formant tracking of the first three formant trajectories in the VTRsynth database that resynthesizes utterances using a stochastic source and formant tracks from the VTR database. Error is only computed over speech-labeled frames.}
\begin{ruledtabular}
\begin{tabular}{cccc}
\textbf{Formant} & \textbf{KARMA} & \textbf{WaveSurfer} & \textbf{Praat} \\
\hline
$f_1$ 			& $29$~Hz & $37$~Hz &  $58$~Hz \\
$f_2$ 			& $53$~Hz & $60$~Hz & $123$~Hz \\
$f_3$ 			& $64$~Hz & $54$~Hz & $130$~Hz \\
Overall 		& $48$~Hz & $50$~Hz & $104$~Hz
\end{tabular}
\end{ruledtabular}
\end{table}

\begin{table}[b]
\caption[RMSE of EKS, WaveSurfer, Praat re VTR on speech frames]{\label{tab:rmse3500synthf0}Average RMSE of KARMA, WaveSurfer, and Praat formant tracking of the first three formant trajectories in the VTRsynthf0 database that resynthesizes VTR database utterances using stochastic and periodic sources. Error is only computed over speech-labeled frames.}
\begin{ruledtabular}
\begin{tabular}{cccc}
\textbf{Formant} & \textbf{KARMA} & \textbf{WaveSurfer} & \textbf{Praat} \\
\hline
$f_1$ 			& $44$~Hz & $57$~Hz &  $57$~Hz \\
$f_2$ 			& $53$~Hz & $58$~Hz & $117$~Hz \\
$f_3$ 			& $62$~Hz & $59$~Hz & $111$~Hz \\
Overall  		& $53$~Hz & $58$~Hz &  $95$~Hz
\end{tabular}
\end{ruledtabular}
\end{table}

Table~\ref{tab:rmse3500synth} and Table~\ref{tab:rmse3500synthf0} display performance on the VTRsynth and VTRsynthf0 databases, respectively, of the three tested algorithms with settings as described in the previous section. The proposed KARMA approach compares favorably to WaveSurfer and Praat. The similar error of KARMA and WaveSurfer validates the use of ARMA cepstral coefficients as observations in place of ARMA spectral coefficients.

\subsection{Antiformant tracking}

The KARMA approach to formant and antiformant tracking is illustrated in this section. Synthesized and real speech examples are presented to determine the ability of the ARMA-derived cepstral coefficients to capture pole and zero information.

\subsubsection{Synthesized waveform}

In the synthesized case, a speech-like waveform /n\textscripta n/ is generated with varying frame-by-frame formant and antiformant characteristics and a periodic source excitation as was implemented for the VTRsynthf0 database. The /n\textscripta n/ waveform is synthesized at $f_s=10$~kHz using $75$ $100$~ms frames with $50$~\% overlap. Formant frequencies (bandwidths) of the /n/ phonemes are set to $257$~Hz ($32$~Hz) and $1891$~Hz ($100$~Hz). One antiformant is placed at $1223$~Hz (bandwidth of $52$~Hz) to mimic the location of an alveolar nasal antiformant. Formant frequencies (bandwidths) of /\textscripta/ were set to $850$~Hz ($80$~Hz) and $1500$~Hz ($120$~Hz). A random term with zero mean and standard deviation of $10$~Hz was added to each trajectory to simulate realistic variation.

\begin{figure*}
  \includegraphics{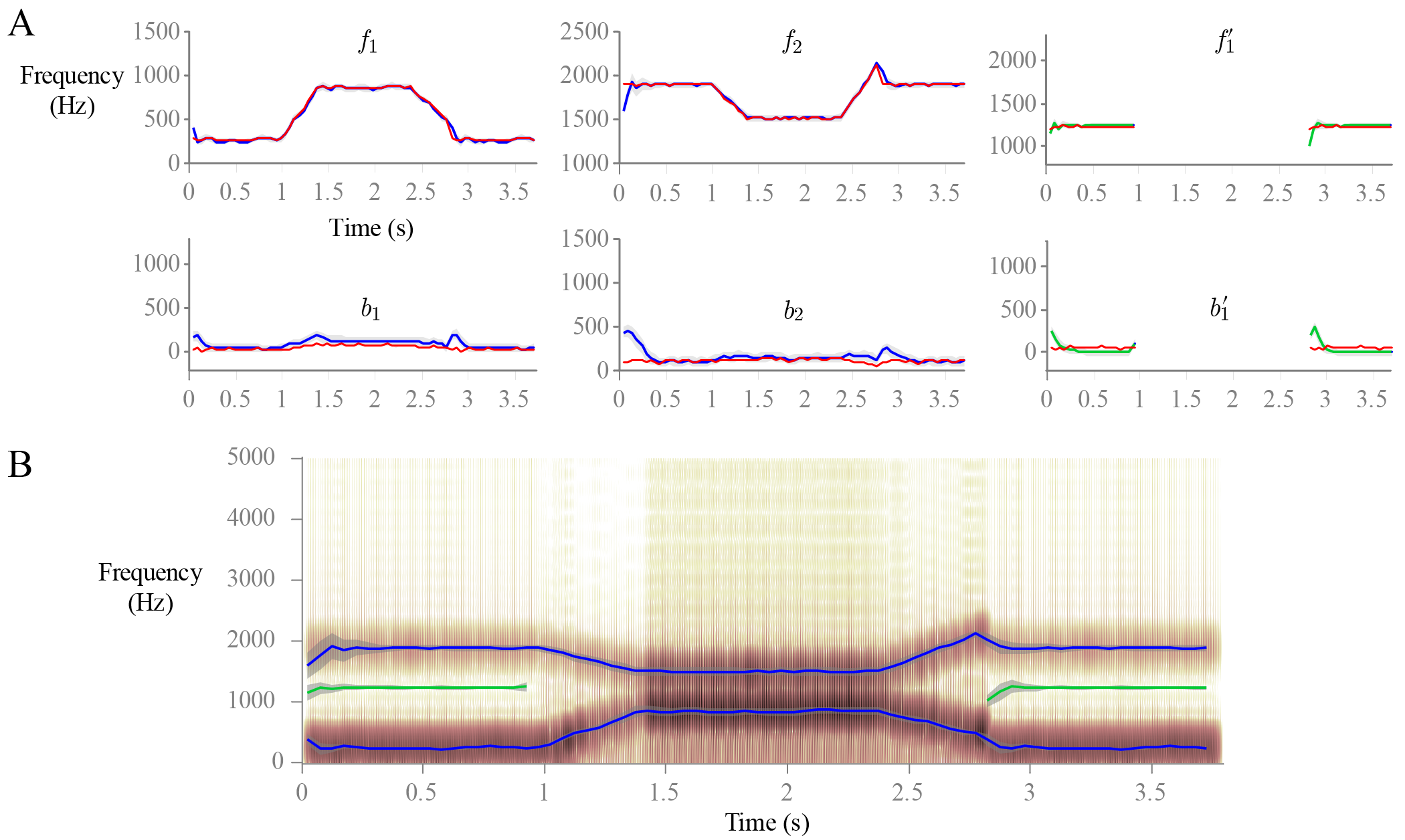}
  \caption{(Color online) Illustration of the output from KARMA for the synthesized utterance /n\textscripta n/. Plots in panel A overlay the true trajectories (red) with the mean estimates (blue for formants, green for antiformants) and uncertainties (gray shading) for each frequency and bandwidth. Panel B plots an alternative display with a wideband spectrogram along with estimated frequency and bandwidth tracks of formants (blue) and antiformants (green). The $3$-dB bandwidths dictate the width of the corresponding frequency tracks.}
  \label{fig:ft_OLA}
\end{figure*}

Fig.~\ref{fig:ft_OLA} shows the results of formant and antiformant tracking using KARMA on the synthesized phoneme string /n\textscripta n/. Two different visualizations are displayed. Fig.~\ref{fig:ft_OLA}A plots point estimates and uncertanties of the center frequency \emph{and} bandwidth trajectories for each frame. Fig.~\ref{fig:ft_OLA}B displays the wideband spectrogram with overlaid center frequency tracks whose width reflects the corresponding $3$-dB bandwidth value. Note that the length of the state vector in the KARMA's state-space model is modified depending on the presence or absence of antiformant energy. Estimated trajectories fit the ground truth values well once initialized values reach a steady state.

\subsubsection{Spoken nasals}

During real speech, a vocal tract configuration consisting of multiple acoustic paths results in the possible existence of both poles and zeros in transfer function $T(z)$ (Eq.~\ref{eq:envelopeSpectrum}). For example, the effects of antiresonances might enter the transfer function of nasalized speech sounds as zeros in $T(z)$. Typically, the frequency of the lowest zero depends on tongue position. For the labial nasal consonant /m/, the frequency of this antiresonance is approximately $1100$~Hz. As the point of closure moves toward the back of the oral cavity---such as for the alveolar and velar nasal consonants---the length of the resonator decreases, and the frequency of this zero increases. The frequency of a second zero is approximately three times the frequency of the lowest zero due to the quarter-wavelength oral cavity configuration.

KARMA performance was evaluated visually on spoken nasal consonants produced with closure at the labial (/m/), alveolar (/n/), and velar (/\textipa{N}/) positions. The extended Kalman smoother was applied using an ARMA($16,4$) model, $f_s=8$~kHz, $20$~ms Hamming windows with $50$~\% overlap, $N=20$ cepstral coefficients, and $\gamma = 0.7$. The frequencies (bandwidths) of the formants were initialized to $500$~Hz ($80$~Hz), $1500$~Hz ($120$~Hz), and $2500$~Hz ($160$~Hz). The frequencies (bandwidths) of the antiformants were initialized to $1000$~Hz ($80$~Hz) and $2000$~Hz ($80$~Hz).

\begin{figure*}
  \includegraphics{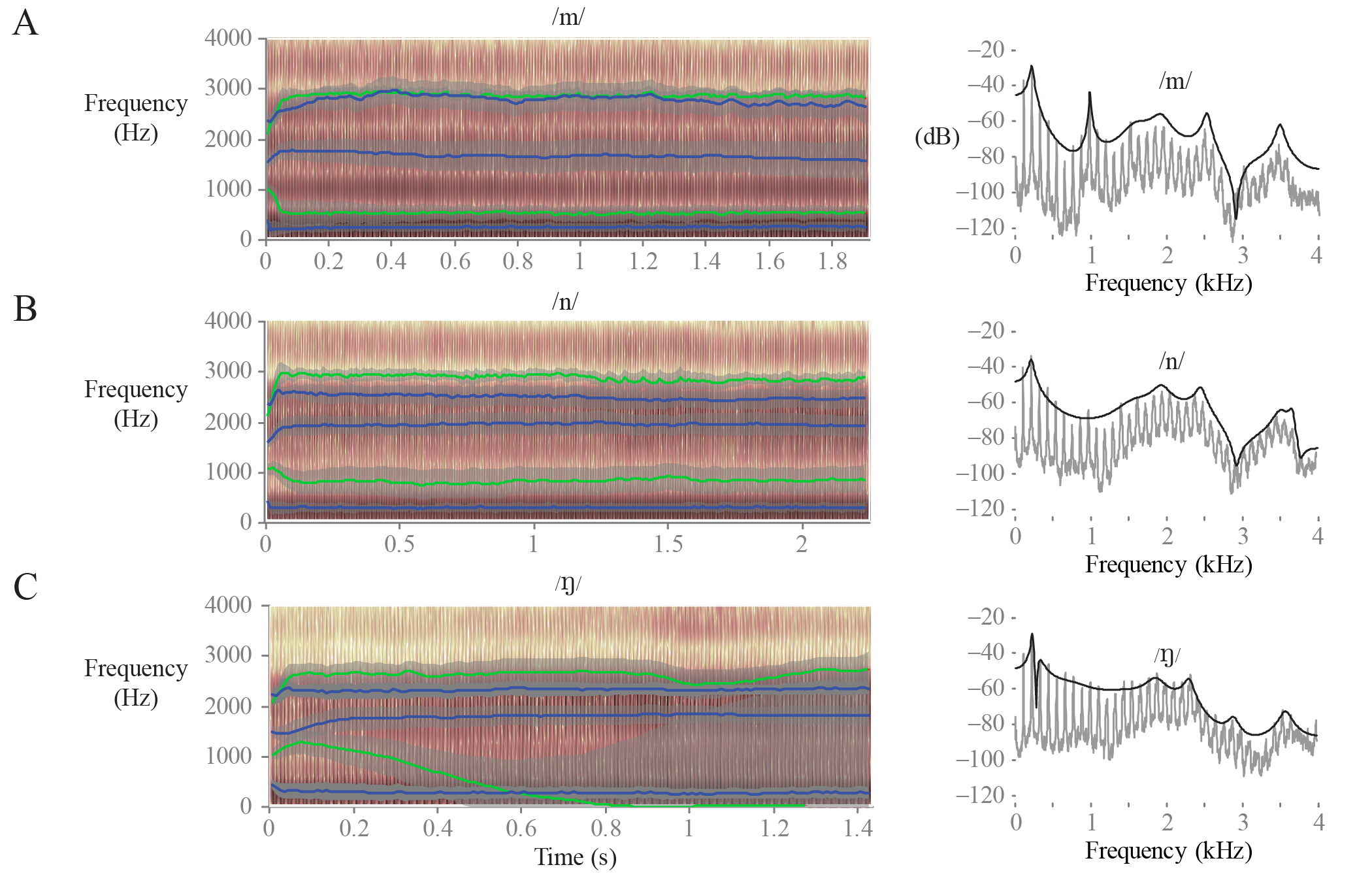}
  \caption{(Color online) KARMA output for three spoken nasals: (A) /m/, (B) /n/, and (C) /\textipa{N}/. On the left, spectrograms overlay the mean estimates (blue for formants, green for antiformants) and uncertainties (gray shading) for each frequency and bandwidth. Plots to the right display the corresponding periodogram (gray) and spectral ARMA model fit (black).}
  \label{fig:ft_EKS_n_m_ng}
\end{figure*}

Figure~\ref{fig:ft_EKS_n_m_ng} displays KARMA outputs (point estimate and uncertainty of frequency tracks) and averaged spectra for the three sustained consonants. The KARMA algorithm takes a few frames to settle to its steady-state estimates. As expected, the frequency of the antiformant increases as the position of closure moves toward the back of the oral cavity. The uncertainty of the first antiformant of /\textipa{N}/ increases significantly, indicating that this antiformant is not well observed in the waveform. Note that the inclusion of zeros greatly improves the ability of the ARMA model to fit the underlying waveform spectra.

\begin{figure*}
  \includegraphics{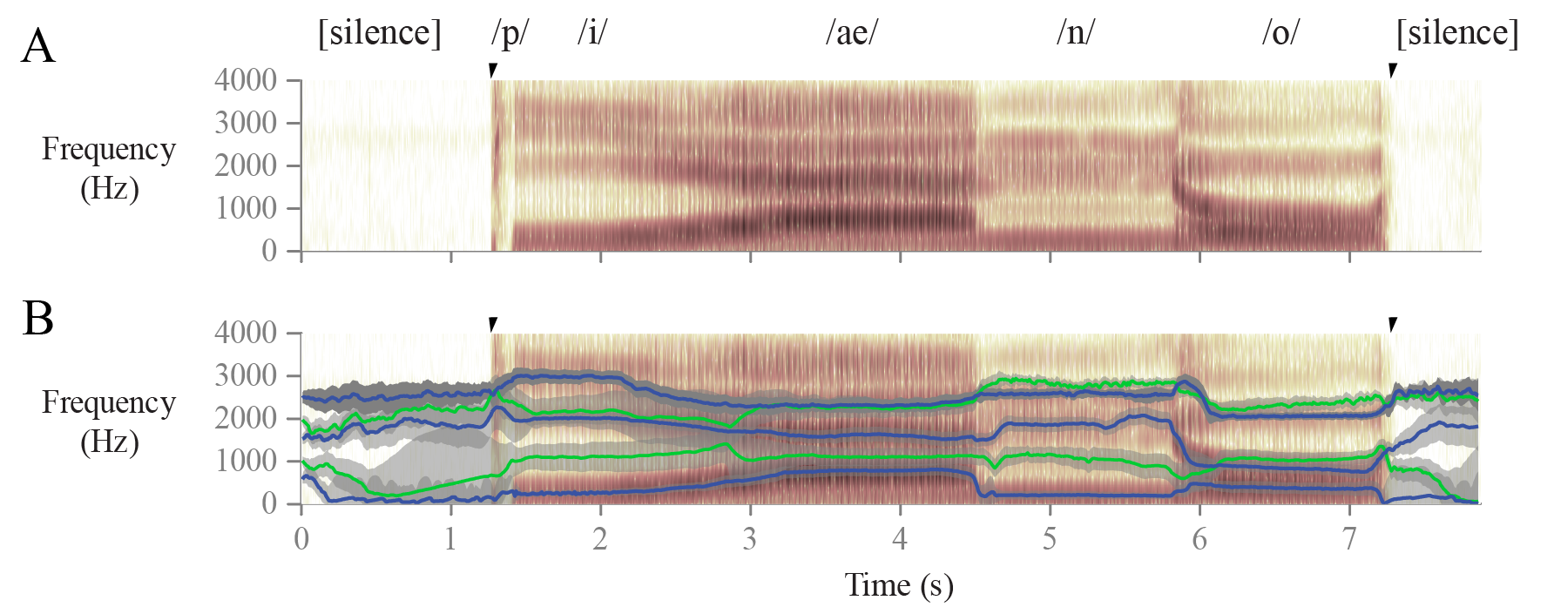}
  \caption{(Color online) KARMA formant and antiformant tracks of utterance by adult male: ``piano.'' Displayed are the (A) wideband spectrogram of the speech waveform and (B) the spectrogram overlaid with formant frequeny estimates (blue), antiformant frequency estimates (green), and uncertainties ($\pm1$ standard deviation) for each track (gray). Arrows indicate beginning and ending of utterance. Note that the increase in uncertainty during silence regions.}
  \label{fig:ft_EKS_piano}
\end{figure*}

Finally, the KARMA tracker was applied to the spoken word ``piano'' to determine if the antiformant tracks would capture any zeros during the nasal phoneme. Figure~\ref{fig:ft_EKS_piano} displays the KARMA formant and antiformant tracks with their associated uncertainties. During the non-nasalized regions, the uncertainty around the point estimates of the antiformant track is large, reflecting the lack of antiresonance information. During the /n/ segment, the uncertainty of the antiformant tracks decreases to reveal observable antiformant information.

\section{Discussion}

In this article, the task of tracking frequencies and bandwidths of formants and antiformants was approached from a statistical point of view. The evolution of parameters was cast in a state-space model to provide access to point estimates and uncertainties of each track. The key relationship was a linearized mapping between cepstral coefficients and formant and antiformant parameters that allowed for the use of the extended family of Kalman inference algorithms.

The VTR database provides an initial benchmark of ``ground truth'' for the first three formant frequency values to which multiple algorithm outputs can be compared. The values in the VTR database, however, should be interpreted with caution because starting values were initially obtained via a first-pass automatic algorithm \cite{Deng2004}. It is unclear how much manual intervention was required and what types of errors were corrected. In particular, VTR tracks do not always overlap high-energy spectral regions. Despite the presence of various labeling errors in the VTR database, it is still useful to obtain initial performance of formant tracking algorithms on real speech.

In the current framework, cepstral coefficients are derived from the spectral coefficients of the fitted stochastic ARMA model (Section~\ref{sec:intraframeObs}). Source information related to phonation is thus separated from vocal tract resonances by assuming that the source is a white Gaussian noise process. This is not the case in reality, especially for voiced speech, where the source excitation component has its own characteristics in frequency (e.g., spectral slope) and time (e.g., periodicity). This model mismatch has been explored here via VTRsynthf0, though we note that it is also possible to incorporate more sophisticated source modeling through the use flexible basis functions such as wavelets \citep{MehtaICASSP2011}.

An alternative approach to ARMA modeling is to compute the nonparametric (real) cepstrum directly from the speech samples. Based on the convolutional model of speech, low-quefrency cepstral coefficients are largely linked to vocal tract information up to about $5$~ms \citep{Childers1977a}, depending on the fundamental frequency. Skipping the $0$th cepstral coefficient that quantifies the overall spectral level, this would translate to including up to the first $35$ cepstral coefficients in the observation vector $\bm{y}_t$ in the state-space framework (Eq.~\ref{eq:stateSpace}). Although coefficients from the real cepstrum do not strictly adhere to the observation model derived in Eq.~\ref{eq:fb2cp}, the approximate separation of source and filter in the nonparametric cepstral domain makes this approach viable.

\begin{figure*}
  \includegraphics{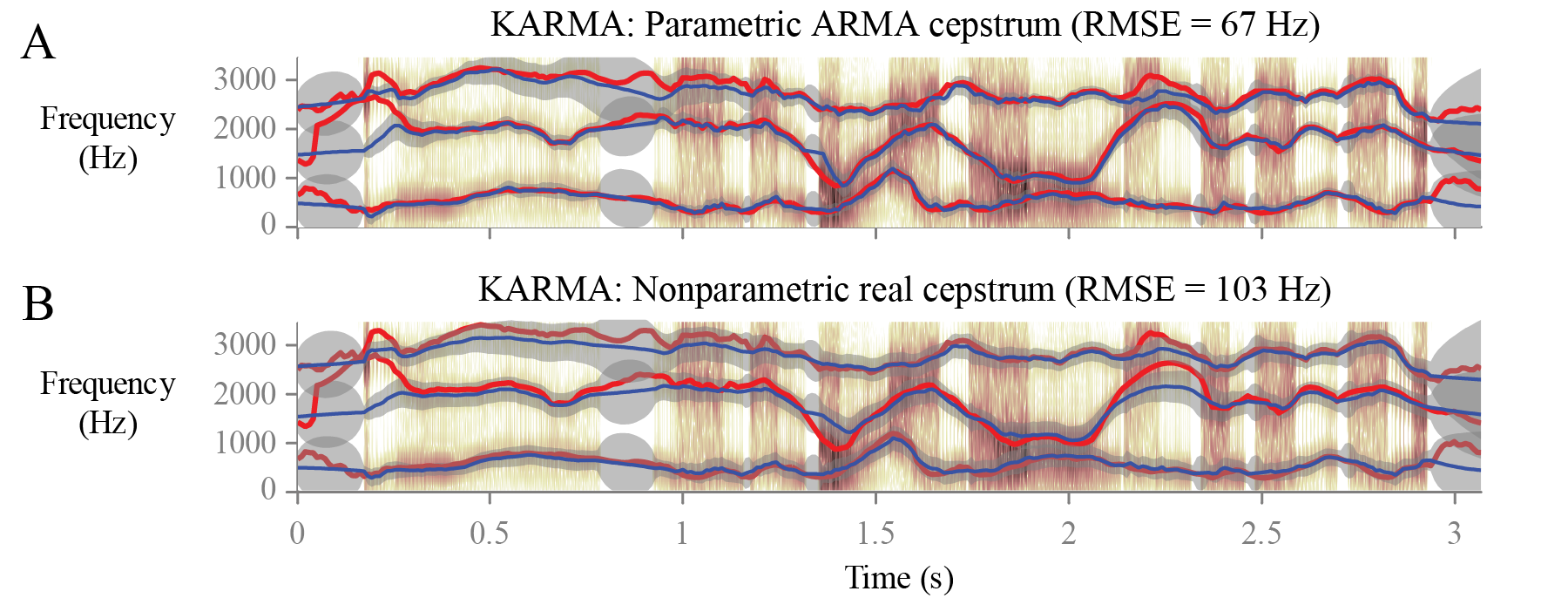}
  \caption{(Color online) KARMA formant tracks using observations from (A) parametric ARMA cepstrum and (B) nonparametric real cepstrum for VTRsynthf0 utterance 1: ``Even then, if she took one step forward, he could catch her.'' Reference trajectories from the VTR database are shown in red with the outputs of KARMA in blue. KARMA uncertainties ($\pm1$ standard deviation) are shown as gray shading. Reported root-mean-square error (RMSE) averages across $3$~formants conditioned on the presence of speech energy for each frame.}
  \label{fig:ft_realCep}
\end{figure*}

Figure~\ref{fig:ft_realCep} illustrates the output of the algorithm using the first $15$ coefficients of the real cepstrum as observations. Interestingly the performance of the nonparametric cepstrum is comparably to that of the parametric ARMA cepstrum, espeically for the first formant frequency. Most of the error stems from underestimating the second and third formant frequencies. Advantages to using the nonparametric cepstrum include computational efficiency and freedom from ARMA model constraints.

The capability of automated methods to track the third formant strongly depends on the resampling frequency, which controls the amount of energy in the spectrum at higher frequencies. For example, if the signal were resampled to $10$~kHz, a given algorithm might erroneously track the third formany frequency through spectral regions typically ascribed to the fourth formant. Traditional formant tracking algorithms have access to multiple candidate frequencies, which are constantly resorted so that $f_4 > f_3 > f_2 > f_1$. In the proposed statistical approach, the ordering of formant indices is inherent in the mapping of formants to cepstral coefficients (\ref{eq:fb2cp}), and further empirical study of this formants-to-cepstrum mapping can be expected to lead to improved methods when there are additional resonances present in the speech bandwidth. Further analysis of ``noise'' in the estimated ARMA cepstrum can also be expected to improve overall robustness in the presence of various sources of uncertainty \citep{Tourneret1995}.

Overall, the proposed KARMA approach compares favorably with WaveSurfer and Praat in terms of root-mean-square error. RMSE, however, is only one selected error metric, which must be validated by observing how well raw trajectories behave. The proposed KARMA tracker yields smoother outputs as a result and offers parameters that allow the user to tune the performance of the algorithm in a statistically principled manner. Such well behaved trajectories may be particularly desirable for the resynthesis of perceptually natural speech.

Though considerably more complex and more sensitive to model assumptions, a time-varying autoregressive moving average (TV-ARMA) model has been previously proposed for formant and antiformant tracking \citep{Toyoshima1991} with little follow-up investigation. In their study, Toyoshima \emph{et al.} used an extended Kalman filter to solve for ARMA spectral coefficients at each speech sample. One real zero and one real pole were included to model changes in gross spectral shape over time. While a frame-based approach (as taken in KARMA) appears to yield more salient parameters at a lower computational cost, future work could consider this as well as alternative time-varying approaches \citep{Rudoy2011}.

As a final observation, antiformant tracking remains a challenging task in speech analysis. Antiresonances are typically less strong than their resonant counterparts during nasalized phonation, and the estimation of subglottal resonances continues to rely on empirical relationships rather than direct acoustic observation \citep{Arsikere2011}. Nevertheless, the proposed approach allows the user the option of tracking antiformants during select speech regions of interest. Potential improvements here include the use of formal statistics tests for detecting the presence of zeros within a frame prior to tracking them.

\section{Conclusions}
This article has presented KARMA, a Kalman-based autoregressive moving average modeling approach to formant and antiformant tracking. The contributions of this work are twofold.  The first is methodological, with improvements to the Kalman-based AR approach of \cite{Deng2007} and extensions to enable antiformant frequency and bandwidth tracking in a KARMA framework. The second is empirical, with visual and quantitative error analysis of the KARMA algorithm demonstrating improvements over two standard speech processing tools, WaveSurfer \citep{Wavesurfer185} and Praat \citep{Praat}.

It is expected that additional improvements will come with better understanding of precisely how formant information is captured through this class of nonlinear ARMA (or nonparametric) cepstral coefficient models. As noted, antiformant tracking remains challenging, although it has been shown here that appropriate results can be obtained for selected cases exhibiting antiresonances. The demonstrated effectiveness of this approach, coupled with its ability to capture uncertainty in the frequency and bandwidth estimates, yields a statistically principled tool appropriate for use in clinical and other applications where it is desired, for example, to quantitatively assess acoustic features such as nasality, subglottal resonances, and coarticulation.


\appendix*

\section{Derivation of cepstral coefficients from the ARMA spectrum}
\label{sec:arma2lpcc}

Assume an ARMA process with the minimum-phase rational transfer function
\begin{equation}
		\label{eq:armaSpectrumA}
    T(z) \triangleq \frac{B(z)}{A(z)} = \frac{1+\sum_{j=1}^{q}b_jz^{-j}}{1-\sum_{i=1}^{p}a_iz^{-i}},
\end{equation}
which in turn implies a right-sided complex cepstrum. For the moment, assume $b_j=0$ for $1 \leq j \leq q$ to initially derive the all-pole LPC cepstrum whose $\mathcal{Z}$-transform is denoted by $C(z)$:
\begin{equation}
    \label{eq:cepZ}
    C(z) \triangleq \log T(z) = \sum_{n = 0}^\infty c_n z^{-n} \text{,}
\end{equation}
where
\begin{equation*}
    c_n = \frac{1}{2\pi }\oint_{z = e^{iw}} \left ( \log T(z) \right ) z^{n-1} dz
\end{equation*}
is the $n$th coefficient of the LPC cepstrum.

Using the chain rule, $\frac{d}{dz^{-1}} T(z)$ can be obtained independently from Eq.~\eqref{eq:armaSpectrumA} or Eq.~\eqref{eq:cepZ}, yielding the relation
\begin{equation*}
    \frac{dC(z)}{dz^{-1}} = \frac{1}{T(z)} \frac{dT(z)}{dz^{-1}} = \frac{\sum_{i=1}^{p}i a_i z^{-i+1} }{1-\sum_{i=1}^{p}a_iz^{-i}} \text{,}
\end{equation*}
which implies that
\begin{equation*}
    \frac{\sum_{i=1}^p i a_i z^{-i+1} }{1-\sum_{i=1}^{p}a_iz^{-i}} = \sum_{n = 0}^\infty c_n \frac{d}{dz^{-1}} \left ( z^{-n} \right ) =\sum_{n = 0}^\infty n c_n z^{-n+1} \text{.}
\end{equation*}
Rearranging the terms above, we obtain
\begin{equation}
    \label{eq:lpcc1}
    \sum_{n = 0}^\infty n c_n z^{-n+1} = \sum_{i=1}^{p}i a_i z^{-i+1} + \sum_{i=1}^{p}a_iz^{-i}\sum_{n = 0}^\infty n c_n z^{-n+1} \text{.}
\end{equation}
Using Eq.~\eqref{eq:lpcc1}, we can match the coefficients of terms on both sides with equal exponents. In the constant-coefficient case (associated to $z^0$), we have $c_1 = a_1 $. For $1 < n \leq p$, we obtain
\begin{equation*}
    c_n = a_n + \sum_{i=1}^{n-1} \frac{n-i}{n}a_i c_{n-i} = a_n + \sum_{i=1}^{n-1} \left (1 - \frac{i}{n} \right ) a_i c_{n-i} \text{.}
\end{equation*}
On the other hand, if $n > p$, then Eq.~\eqref{eq:lpcc1} implies that
\begin{equation*}
    c_n = \sum_{i=n-p}^{n-1} \frac{n-i}{n}a_i c_{n-i} = \sum_{i=1}^{n-1} \left (1 - \frac{i}{n} \right ) a_i c_{n-i} \text{.}
\end{equation*}
In summary, we have obtained the following relationship between the prediction polynomial coefficients and the complex cepstrum:
\begin{equation*}
    \label{eq:ar2cep1}
    c_n = \begin{cases}
        a_1 & \text{if} \quad n = 1 \\
        a_n + \sum_{i=1}^{n-1} \left(\frac{n-i}{n}\right) a_i c_{n-i} & \text{if} \quad 1 < n \leq p \\
        \sum_{i=n-p}^{n-1} \left(\frac{n-i}{n}\right) a_i c_{n-i}  & \text{if} \quad p < n \text{.}
    \end{cases}
\end{equation*}
Reversing the roles of $i$ and $(n-i)$ yields the all-pole version in Eq.~\eqref{eq:arma2lpccP}.

To allow for nonzero $b_j$ coefficients in Eq.~\eqref{eq:armaSpectrumA}, we obtain the ARMA cepstral coefficients $C_n$ by separating contributions from the numerator and denominator of Eq.~\eqref{eq:armaSpectrumA} as follows:
\begin{align*}
    C_n &= \mathcal{Z}^{-1} \log T(z) \\
    	  &= \mathcal{Z}^{-1} \log \frac{1}{A(z)} - \mathcal{Z}^{-1} \log \frac{1}{B(z)} \\
				&= c_n - c'_n \text{,}
\end{align*}
yielding the respective pole and zero recursions of Eqs.~\eqref{eq:arma2lpcc}.

\bibliography{formantTracking_JASAms_v39}

\begin{thebibliography}{35}
\newcommand{\enquote}[1]{``#1''}
\expandafter\ifx\csname natexlab\endcsname\relax\def\natexlab#1{#1}\fi
\expandafter\ifx\csname url\endcsname\relax
  \def\url#1{\texttt{#1}}\fi
\expandafter\ifx\csname urlprefix\endcsname\relax\def\urlprefix{URL }\fi
\providecommand{\bibinfo}[2]{#2}
\providecommand{\noopsort}[1]{}
\providecommand{\switchargs}[2]{#2#1}

\bibitem[{Arsikere \emph{et~al.}(2011)Arsikere, Lulich, and
  Alwan}]{Arsikere2011}
\bibinfo{author}{Arsikere, H.}, \bibinfo{author}{Lulich, S.~M.}, and
  \bibinfo{author}{Alwan, A.} (\textbf{\bibinfo{year}{2011}}).
  \enquote{\bibinfo{title}{Automatic estimation of the first subglottal
  resonance}}, \bibinfo{journal}{J. Acoust. Soc. Am.}
  \textbf{\bibinfo{volume}{129}}, \bibinfo{pages}{EL197--EL203}.

\bibitem[{Atal and Hanauer(1971)}]{Atal1971}
\bibinfo{author}{Atal, B.~S.} and \bibinfo{author}{Hanauer, S.~L.}
  (\textbf{\bibinfo{year}{1971}}). \enquote{\bibinfo{title}{Speech analysis and
  synthesis by linear prediction of the speech wave}}, \bibinfo{journal}{J.
  Acoust. Soc. Am.} \textbf{\bibinfo{volume}{50}}, \bibinfo{pages}{637--655}.

\bibitem[{Atal and Schroeder(1978)}]{Atal1978}
\bibinfo{author}{Atal, B.~S.} and \bibinfo{author}{Schroeder, M.~R.}
  (\textbf{\bibinfo{year}{1978}}). \enquote{\bibinfo{title}{Linear prediction
  analysis of speech based on a pole-zero representation}},
  \bibinfo{journal}{J. Acoust. Soc. Am.} \textbf{\bibinfo{volume}{64}},
  \bibinfo{pages}{1310--1318}.

\bibitem[{Boersma and Weenink(2009)}]{Praat}
\bibinfo{author}{Boersma, P.} and \bibinfo{author}{Weenink, D.}
  (\textbf{\bibinfo{year}{2009}}). \enquote{\bibinfo{title}{Praat: Doing
  phonetics by computer}}, \bibinfo{journal}{version 5.1.40}
  \bibinfo{pages}{retrieved from www.praat.org 13 July 2009}.

\bibitem[{Broad and Clermont(1989)}]{Broad1989}
\bibinfo{author}{Broad, D.~J.} and \bibinfo{author}{Clermont, F.}
  (\textbf{\bibinfo{year}{1989}}). \enquote{\bibinfo{title}{Formant estimation
  by linear transformation of the {LPC} cepstrum}}, \bibinfo{journal}{J.
  Acoust. Soc. Am.} \textbf{\bibinfo{volume}{86}}, \bibinfo{pages}{2013--2017}.

\bibitem[{Childers \emph{et~al.}(1977)Childers, Skinner, and
  Kemerait}]{Childers1977a}
\bibinfo{author}{Childers, D.~G.}, \bibinfo{author}{Skinner, D.~P.}, and
  \bibinfo{author}{Kemerait, R.~C.} (\textbf{\bibinfo{year}{1977}}).
  \enquote{\bibinfo{title}{The cepstrum: A guide to processing}},
  \bibinfo{journal}{Proc. IEEE} \textbf{\bibinfo{volume}{65}},
  \bibinfo{pages}{1428--1443}.

\bibitem[{Christensen \emph{et~al.}(1976)Christensen, Strong, and
  Palmer}]{Christensen1976}
\bibinfo{author}{Christensen, R.}, \bibinfo{author}{Strong, W.}, and
  \bibinfo{author}{Palmer, E.} (\textbf{\bibinfo{year}{1976}}).
  \enquote{\bibinfo{title}{A comparison of three methods of extracting
  resonance information from predictor-coefficient coded speech}},
  \bibinfo{journal}{IEEE Trans. Acoust.} \textbf{\bibinfo{volume}{24}},
  \bibinfo{pages}{8--14}.

\bibitem[{Deng \emph{et~al.}(2006{\natexlab{a}})Deng, Acero, and
  Bazzi}]{Deng2006}
\bibinfo{author}{Deng, L.}, \bibinfo{author}{Acero, A.}, and
  \bibinfo{author}{Bazzi, I.} (\textbf{\bibinfo{year}{2006}}{\natexlab{a}}).
  \enquote{\bibinfo{title}{Tracking vocal tract resonances using a quantized
  nonlinear function embedded in a temporal constraint}},
  \bibinfo{journal}{IEEE Trans. Audio Speech Lang. Processing}
  \textbf{\bibinfo{volume}{14}}, \bibinfo{pages}{425--434}.

\bibitem[{Deng \emph{et~al.}(2006{\natexlab{b}})Deng, Cui, Pruvenok, Huang,
  Momen, Chen, and Alwan}]{Deng2006a}
\bibinfo{author}{Deng, L.}, \bibinfo{author}{Cui, X.},
  \bibinfo{author}{Pruvenok, R.}, \bibinfo{author}{Huang, J.},
  \bibinfo{author}{Momen, S.}, \bibinfo{author}{Chen, Y.}, and
  \bibinfo{author}{Alwan, A.} (\textbf{\bibinfo{year}{2006}}{\natexlab{b}}).
  \enquote{\bibinfo{title}{A database of vocal tract resonance trajectories for
  research in speech processing}}, \bibinfo{journal}{Proc. IEEE Int. Conf.
  Acoust. Speech Signal Process.} \textbf{\bibinfo{volume}{1}},
  \bibinfo{pages}{369--372}.

\bibitem[{Deng \emph{et~al.}(2004)Deng, Lee, Attias, and Acero}]{Deng2004}
\bibinfo{author}{Deng, L.}, \bibinfo{author}{Lee, L.~J.},
  \bibinfo{author}{Attias, H.}, and \bibinfo{author}{Acero, A.}
  (\textbf{\bibinfo{year}{2004}}). \enquote{\bibinfo{title}{A structured speech
  model with continuous hidden dynamics and prediction-residual training for
  tracking vocal tract resonances}}, \bibinfo{journal}{IEEE International
  Conference on Acoustics, Speech, and Signal Processing}
  \textbf{\bibinfo{volume}{1}}, \bibinfo{pages}{I--557--60}.

\bibitem[{Deng \emph{et~al.}(2007)Deng, Lee, Attias, and Acero}]{Deng2007}
\bibinfo{author}{Deng, L.}, \bibinfo{author}{Lee, L.~J.},
  \bibinfo{author}{Attias, H.}, and \bibinfo{author}{Acero, A.}
  (\textbf{\bibinfo{year}{2007}}). \enquote{\bibinfo{title}{Adaptive {K}alman
  filtering and smoothing for tracking vocal tract resonances using a
  continuous-valued hidden dynamic model}}, \bibinfo{journal}{IEEE Trans. Audio
  Speech Lang. Processing} \textbf{\bibinfo{volume}{15}},
  \bibinfo{pages}{13--23}.

\bibitem[{Fulop(2010)}]{Fulop2010}
\bibinfo{author}{Fulop, S.~A.} (\textbf{\bibinfo{year}{2010}}).
  \enquote{\bibinfo{title}{Accuracy of formant measurement for synthesized
  vowels using the reassigned spectrogram and comparison with linear
  prediction}}, \bibinfo{journal}{J. Acoust. Soc. Am.}
  \textbf{\bibinfo{volume}{127}}, \bibinfo{pages}{2114--2117}.

\bibitem[{Garofolo \emph{et~al.}(1993)Garofolo, Lamel, Fisher, Fiscus, Pallett,
  Dahlgren, and Zue}]{TIMIT}
\bibinfo{author}{Garofolo, J.~S.}, \bibinfo{author}{Lamel, L.},
  \bibinfo{author}{Fisher, W.}, \bibinfo{author}{Fiscus, J.},
  \bibinfo{author}{Pallett, D.}, \bibinfo{author}{Dahlgren, N.}, and
  \bibinfo{author}{Zue, V.} (\textbf{\bibinfo{year}{1993}}).
  \emph{\bibinfo{title}{{TIMIT} Acoustic-Phonetic Continuous Speech Corpus}}
  (\bibinfo{publisher}{Linguistic Data Consortium},
  \bibinfo{address}{Philadelphia, PA}).

\bibitem[{Hamilton(1994)}]{Hamilton1994}
\bibinfo{author}{Hamilton, J.~D.} (\textbf{\bibinfo{year}{1994}}).
  \emph{\bibinfo{title}{Time Series Analysis}} (\bibinfo{publisher}{Princeton
  University}).

\bibitem[{Julier and Uhlmann(1997)}]{Julier1997}
\bibinfo{author}{Julier, S.} and \bibinfo{author}{Uhlmann, J.}
  (\textbf{\bibinfo{year}{1997}}). \enquote{\bibinfo{title}{A new extension of
  the {K}alman filter to nonlinear systems}}, \bibinfo{journal}{Proceedings of
  AeroSense: The 11th International Symposium on Aerospace/Defense Sensing,
  Simulation, and Controls} \textbf{\bibinfo{volume}{3}}, \bibinfo{pages}{26}.

\bibitem[{Kalman(1960)}]{Kalman1960}
\bibinfo{author}{Kalman, R.~E.} (\textbf{\bibinfo{year}{1960}}).
  \enquote{\bibinfo{title}{A new approach to linear filtering and prediction
  problems}}, \bibinfo{journal}{Trans. ASME J. Basic Eng.}
  \textbf{\bibinfo{volume}{82}}, \bibinfo{pages}{35--45}.

\bibitem[{Klatt(1980)}]{Klatt1980}
\bibinfo{author}{Klatt, D.~H.} (\textbf{\bibinfo{year}{1980}}).
  \enquote{\bibinfo{title}{Software for a cascade/parallel formant
  synthesizer}}, \bibinfo{journal}{J. Acoust. Soc. Am.}
  \textbf{\bibinfo{volume}{67}}, \bibinfo{pages}{971--995}.

\bibitem[{Kopec(1986)}]{Kopec1986}
\bibinfo{author}{Kopec, G.} (\textbf{\bibinfo{year}{1986}}).
  \enquote{\bibinfo{title}{Formant tracking using hidden markov models and
  vector quantization}}, \bibinfo{journal}{IEEE Trans. Acoust.}
  \textbf{\bibinfo{volume}{34}}, \bibinfo{pages}{709--729}.

\bibitem[{Ljung(1999)}]{Ljung1999}
\bibinfo{author}{Ljung, L.} (\textbf{\bibinfo{year}{1999}}).
  \emph{\bibinfo{title}{System {Identification}}} (\bibinfo{publisher}{Upper
  Saddle River, NJ: Prentice-Hall}).

\bibitem[{Marelli and Balazs(2010)}]{Marelli2010}
\bibinfo{author}{Marelli, D.} and \bibinfo{author}{Balazs, P.}
  (\textbf{\bibinfo{year}{2010}}). \enquote{\bibinfo{title}{On pole-zero model
  estimation methods minimizing a logarithmic criterion for speech analysis}},
  \bibinfo{journal}{IEEE Trans. Audio Speech Lang. Processing}
  \textbf{\bibinfo{volume}{18}}, \bibinfo{pages}{237--248}.

\bibitem[{McCandless(1974)}]{McCandless1974}
\bibinfo{author}{McCandless, S.} (\textbf{\bibinfo{year}{1974}}).
  \enquote{\bibinfo{title}{An algorithm for automatic formant extraction using
  linear prediction spectra}}, \bibinfo{journal}{IEEE Trans. Acoust.}
  \textbf{\bibinfo{volume}{22}}, \bibinfo{pages}{134--141}.

\bibitem[{Mehta \emph{et~al.}(2011)Mehta, Rudoy, and Wolfe}]{MehtaICASSP2011}
\bibinfo{author}{Mehta, D.~D.}, \bibinfo{author}{Rudoy, D.}, and
  \bibinfo{author}{Wolfe, P.~J.} (\textbf{\bibinfo{year}{2011}}).
  \enquote{\bibinfo{title}{Joint source-filter modeling using flexible basis
  functions}}, \bibinfo{journal}{IEEE Int. Conf. Acoust. Speech Signal
  Processing} .

\bibitem[{Miyanaga \emph{et~al.}(1986)Miyanaga, Miki, and Nagai}]{Miyanaga1986}
\bibinfo{author}{Miyanaga, Y.}, \bibinfo{author}{Miki, N.}, and
  \bibinfo{author}{Nagai, N.} (\textbf{\bibinfo{year}{1986}}).
  \enquote{\bibinfo{title}{Adaptive identification of a time-varying arma
  speech model}}, \bibinfo{journal}{IEEE Trans. Acoust.}
  \textbf{\bibinfo{volume}{34}}, \bibinfo{pages}{423--433}.

\bibitem[{Nearey(1989)}]{Nearey1989}
\bibinfo{author}{Nearey, T.~M.} (\textbf{\bibinfo{year}{1989}}).
  \enquote{\bibinfo{title}{Static, dynamic, and relational properties in vowel
  perception}}, \bibinfo{journal}{J. Acoust. Soc. Am.}
  \textbf{\bibinfo{volume}{85}}, \bibinfo{pages}{2088--2113}.

\bibitem[{Rigoll(1986)}]{Rigoll1986}
\bibinfo{author}{Rigoll, G.} (\textbf{\bibinfo{year}{1986}}).
  \enquote{\bibinfo{title}{A new algorithm for estimation of formant
  trajectories directly from the speech signal based on an extended
  kalman-filter}}, \bibinfo{journal}{Proc. IEEE Int. Conf. Acoust. Speech
  Signal Process.} \textbf{\bibinfo{volume}{11}}, \bibinfo{pages}{1229--1232}.

\bibitem[{Rosenberg(1971)}]{Rosenberg1971}
\bibinfo{author}{Rosenberg, A.~E.} (\textbf{\bibinfo{year}{1971}}).
  \enquote{\bibinfo{title}{Effect of glottal pulse shape on the quality of
  natural vowels}}, \bibinfo{journal}{J. Acoust. Soc. Am.}
  \textbf{\bibinfo{volume}{49}}, \bibinfo{pages}{583--590}.

\bibitem[{Rudoy \emph{et~al.}(2011)Rudoy, Quatieri, and Wolfe}]{Rudoy2011}
\bibinfo{author}{Rudoy, D.}, \bibinfo{author}{Quatieri, T.}, and
  \bibinfo{author}{Wolfe, P.} (\textbf{\bibinfo{year}{2011}}).
  \enquote{\bibinfo{title}{Time-varying autoregressions in speech: Detection
  theory and applications}}, \bibinfo{journal}{IEEE Trans. Audio Speech Lang.
  Processing} \textbf{\bibinfo{volume}{19}}, \bibinfo{pages}{977--989}.

\bibitem[{Rudoy \emph{et~al.}(2007)Rudoy, Spendley, and Wolfe}]{Rudoy2007}
\bibinfo{author}{Rudoy, D.}, \bibinfo{author}{Spendley, D.~N.}, and
  \bibinfo{author}{Wolfe, P.~J.} (\textbf{\bibinfo{year}{2007}}).
  \enquote{\bibinfo{title}{Conditionally linear {G}aussian models for
  estimating vocal tract resonances}}, \bibinfo{journal}{Proc. INTERSPEECH}
  \bibinfo{pages}{526--529}.

\bibitem[{Schafer and Rabiner(1970)}]{Schafer1970}
\bibinfo{author}{Schafer, R.~W.} and \bibinfo{author}{Rabiner, L.~R.}
  (\textbf{\bibinfo{year}{1970}}). \enquote{\bibinfo{title}{System for
  automatic formant analysis of voiced speech}}, \bibinfo{journal}{J. Acoust.
  Soc. Am.} \textbf{\bibinfo{volume}{47}}, \bibinfo{pages}{634--648}.

\bibitem[{Sj\"olander and Beskow(2005)}]{Wavesurfer185}
\bibinfo{author}{Sj\"olander, K.} and \bibinfo{author}{Beskow, J.}
  (\textbf{\bibinfo{year}{2005}}). \enquote{\bibinfo{title}{{W}ave{S}urfer for
  {W}indows}}, \bibinfo{journal}{version 1.8.5} \bibinfo{pages}{retrieved 1
  November 2005}.

\bibitem[{Steiglitz(1977)}]{Steiglitz1977}
\bibinfo{author}{Steiglitz, K.} (\textbf{\bibinfo{year}{1977}}).
  \enquote{\bibinfo{title}{On the simultaneous estimation of poles and zeros in
  speech analysis}}, \bibinfo{journal}{IEEE Trans. Acoust.}
  \textbf{\bibinfo{volume}{25}}, \bibinfo{pages}{229--234}.

\bibitem[{Tourneret and Lacaze(1995)}]{Tourneret1995}
\bibinfo{author}{Tourneret, J.-Y.} and \bibinfo{author}{Lacaze, B.}
  (\textbf{\bibinfo{year}{1995}}). \enquote{\bibinfo{title}{On the statistics
  of estimated reflection and cepstrum coefficients of an autoregressive
  process}}, \bibinfo{journal}{Signal Processing}
  \textbf{\bibinfo{volume}{43}}, \bibinfo{pages}{253--267}.

\bibitem[{Toyoshima \emph{et~al.}(1991)Toyoshima, Miki, and
  Nagai}]{Toyoshima1991}
\bibinfo{author}{Toyoshima, T.}, \bibinfo{author}{Miki, N.}, and
  \bibinfo{author}{Nagai, N.} (\textbf{\bibinfo{year}{1991}}).
  \enquote{\bibinfo{title}{Adapative formant estimation with compensation for
  gross spectral shape}}, \bibinfo{journal}{Electronics and Communications in
  Japan (Part III: Fundamental Electronic Science)}
  \textbf{\bibinfo{volume}{74}}, \bibinfo{pages}{58--68}.

\bibitem[{Yegnanarayana(1978)}]{Yegnanarayana1978}
\bibinfo{author}{Yegnanarayana, B.} (\textbf{\bibinfo{year}{1978}}).
  \enquote{\bibinfo{title}{Formant extraction from linear-prediction phase
  spectra}}, \bibinfo{journal}{J. Acoust. Soc. Am.}
  \textbf{\bibinfo{volume}{63}}, \bibinfo{pages}{1638--1640}.

\bibitem[{Zheng and Hasegawa-Johnson(2004)}]{Zheng2004}
\bibinfo{author}{Zheng, Y.} and \bibinfo{author}{Hasegawa-Johnson, M.}
  (\textbf{\bibinfo{year}{2004}}). \enquote{\bibinfo{title}{Formant tracking by
  mixture state particle filter}}, \bibinfo{journal}{Proc. IEEE Int. Conf.
  Acoust. Speech Signal Process.} \textbf{\bibinfo{volume}{1}},
  \bibinfo{pages}{565--568}.

\end{thebibliography}

\end{document}